\documentclass[aps,twocolumn]{revtex4}
\usepackage{graphicx}

\begin{document}
\title{Wrinkling of freely floating smectic films}

\author{K. Harth$^{1,2}$, T. Trittel$^1$, K. May$^1$, and R. Stannarius$^1$} 

\affiliation{$^1$Institute of Physics, Otto von Guericke University, 39106 Magdeburg, Germany,
$^{2}$Universiteit Twente, Physics of Fluids and Max Planck Center for Complex Fluid Dynamics,
7500 AE Enschede, The Netherlands}

\begin{abstract}
We demonstrate spontaneous wrinkling as a transient dynamical pattern in thin freely floating smectic liquid-crystalline films.
The peculiarity of such films is that, while flowing liquid-like in the film plane,
they cannot quickly expand in the direction perpendicular to that plane. At short time scales they therefore
behave in two dimensions like quasi-incompressible membranes. Such films can develop a transient
undulation instability or form bulges in response to lateral compression.
Optical experiments with freely floating bubbles on parabolic flights and in ground lab experiments are reported.
The characteristic wavelengths of the wrinkles are in the submillimeter range. We demonstrate the dynamic nature
of the pattern formation mechanism and develop a basic model for the wavelength selection and wrinkle orientation.
\end{abstract}

\maketitle

\section{Introduction}

Wrinkling is a frequent response of soft surfaces or thin sheets to lateral stresses or stress mismatches.
Commonly, it represents a stable or metastable energy minimum of an elastically deformed structure.
In thin films of common fluids, such as free-standing films of soap solutions, or low viscosity liquids on an
elastic substrate, a lateral compression and reduction of the film area can be easily compensated by local
film thickness changes. Thus, these films do never wrinkle or buckle. In fluid films with internal structures
such as smectic free-standing films or lipid membranes, the situation is different. A reduction of the film
extension requires structural reorganizations which may be slow compared to the compression rate, and wrinkles
may form to compensate lateral compressions that occur sufficiently fast.

Owing to their internal molecular layer structure, some smectic phases can form stable freely suspended
films with submicrometer thickness and up to several centimeters in lateral extension. Smectic A and C phases
exhibit fluid-like characteristics with regard to flow in the film plane.
Such films have many features in common with soap films, and they can be prepared, similar to the latter,
in the form of bubbles \citep{Oswald1987,Stannarius1997,Stannarius1998}.
At first glance, freely floating smectic bubbles appear very similar to soap bubbles, but they differ from
the latter qualitatively in some dynamic properties
\citep{May2012,May2014}, as a consequence of the layered molecular arrangement.

\begin{figure}
\center
\includegraphics[width=0.8\columnwidth]{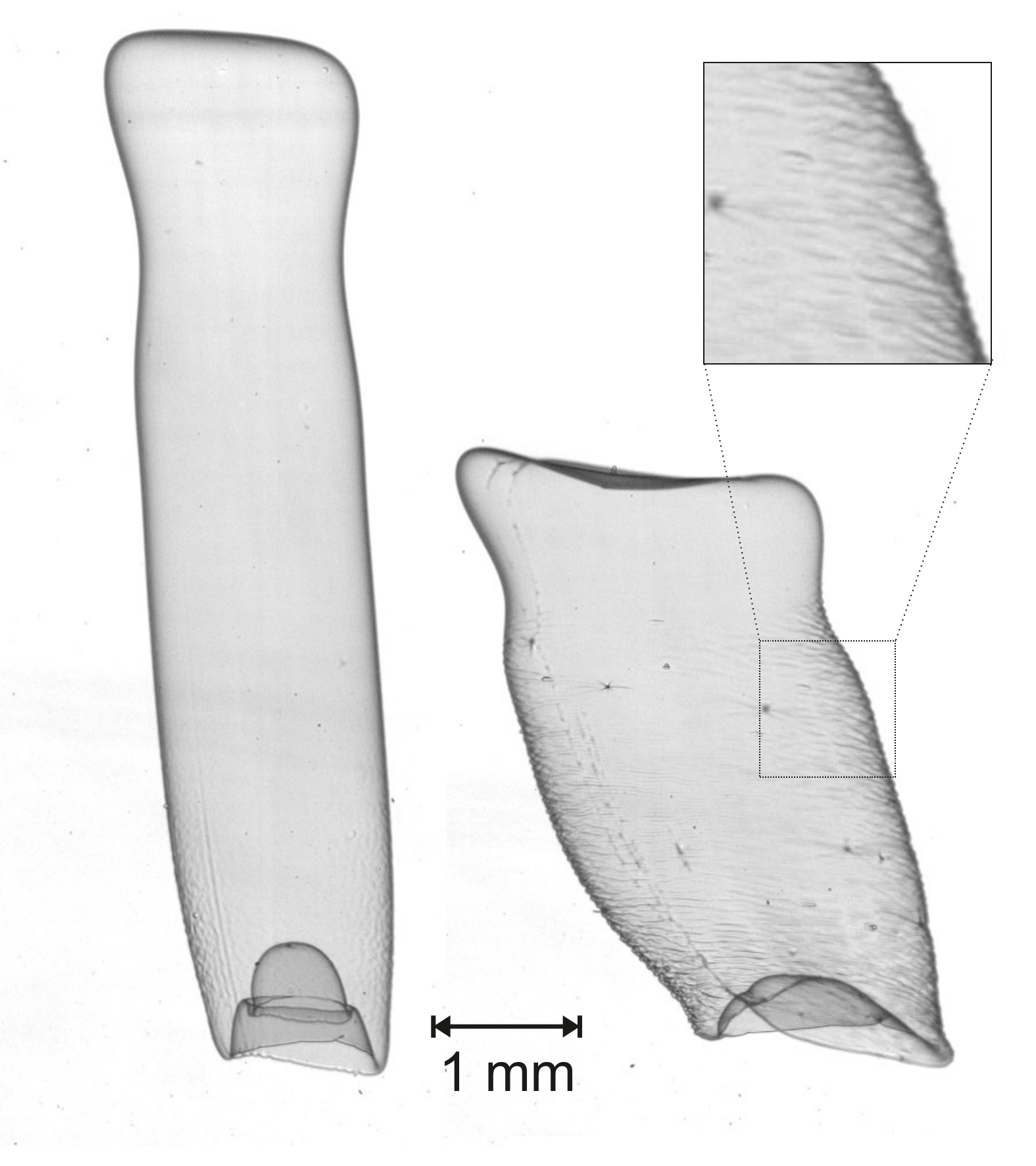}	
\caption{Freely floating smectic bubble with homogeneous film thickness of 50 nm
that is axially compressed by air flow.
Thereby, it develops a wrinkled structure normal to the compression direction. The right hand image was recorded 50 ms after the left one.
The bubble was produced under normal gravity, thus the pinch-off is asymmetric and
the bottom part develops a typical invagination.
}
	\label{fig:1}
\end{figure}

\begin{figure}
\center  \includegraphics[width=0.8\columnwidth]{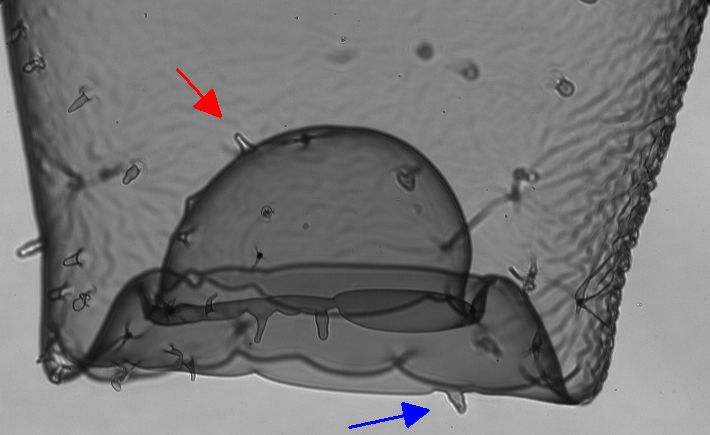}
	\caption{Detail of a freely floating smectic bubble near the invagination at one end, with wrinkles and tubular extrusions that form during shape relaxation. Extrusions of the bubble can point inwards or outwards (see arrows). Wrinkles appear, for example, when the bubble is rapidly compressed in one or two directions. The image width is approximately 1~mm.}
	\label{fig:1a}
\end{figure}

Any reduction of the local film surface requires the growth of the film thickness to conserve the volume
of the smectic material. The related reorganization of smectic layers is achieved in the films by the formation
of so-called  islands, viz. spots with additional molecular layers on the film that are enclosed by dislocation
arrays \citep{Oswald1987,May2014,May2012,Daehmlow2018}.
Such a structural rearrangement represents a comparably slow process, it requires up to several hundred
milliseconds.
Thus, on short time scales these free-standing films are not able to respond to external forcing by thickness adaption.
They retain a roughly constant surface area. When exposed to fast geometrical shape transformations,
they behave more similar to thin elastic sheets than like fluid films.
An effective demonstration of a wrinkle formation should be achieved by the rapid reduction of the area of
a rectangular frame supporting such films. However, this is technically very challenging.
As we have shown, very rapid shape changes can be observed in freely floating smectic films
and bubbles \citep{May2014}. We describe in this manuscript that
a fast and efficient response of the smectic film to such changes is the formation of bulges and
tubuli or wrinkles. Thereby, the film surface remains
roughly constant, yet the local deformations result in a reduction of the effective lateral extension of the film
(shortly referred to as 'film area' in the following).
In the present paper, we characterize this unique dynamic wrinkling and we describe the spontaneous formation of
bulges and tubuli (as shown in Figs.~\ref{fig:1},\ref{fig:1a}). A model is derived that gives a qualitative explanation of the main features of the instability.

We performed most of the experiments in microgravity
during parabolic flights, a few studies were carried out in the ground lab at
normal gravity.

\section{Wrinkling mechanisms}

Before describing the experiment and data analysis, we give a short introduction into wrinkling
mechanisms of thin elastic or fluid membranes:
Wrinkles in soft matter systems can be of very different types, static and dynamic.
Examples of static structures are skin wrinkles \citep{Cerda2003}, drying fruit and vegetable peels, or stiff
surface layers on shrinking elastic supports. They are quasi-static responses to stress balances.
The latter also applies to curtains or drapes, where the spontaneous crinkles
reflect a compromise between gravitation energy and bending stiffness. Those structures represent
stable or metastable energy minima.

Mechanically induced static layer undulations have also been discovered in smectic liquid crystals confined to thin sandwich cells with fixed boundary conditions. One example is the classical Helfrich-Hurault instability \citep{Helfrich1970,Bevilacqua2005}. It can occur in thin cells under dilation strain when the cell thickness is increased by external forcing.
Layer undulations relieve smectic layer dilation, the elastic energy necessary for layer thickness changes competes with the elastic energy of director distortions induced by layer curvature. Wavelengths of those patterns are typically in the nanometer range. A similar instability can be produced in a rigid cell by means of electric or magnetic fields.
These are conservative equilibrium patterns as well, representing free energy minima.

On the other hand, dynamic wrinkles can form, e. g. during a collapse of fluid bubbles
in a gaseous environment. When a hemispherical bubble of highly viscous material sessile above a bulk bath ruptures,
it can develop well-ordered wrinkles. The driving forces are gravitation and lateral confinement by the bubble
geometry \citep{Debregeas1998,daSilveira2000,Aumaitre2013}. The formation of wrinkles of fluid interfaces have also been described
during the impact of droplets onto a viscous liquid \citep{Li2017} and for droplets rising in a liquid \citep{Uemura2010}.
A related phenomenon is the buckling of viscous fluid filaments \citep{Merrer2012,Salili2016}.

The structures described here are also of dynamical origin. They belong to the
class of transient pattern forming instabilities. The smectic films undergo a rapid change from an initially
flat to an undulated shape. Thereby, a broad band of modes with different wavelengths becomes
unstable, and the mode with the fastest growth rate dominates the periodicity of
the pattern. We note that the area of the film-air interface does not change noticeably during this wrinkling process,
we can disregard any forces related to surface tension. The thickness of the films is typically
in the range of a few dozen nanometers, much smaller than the observed wavelengths. Therefore we can
treat the film as a quasi-twodimensional (2D) fluid sheet, surrounded by air on both sides.
Because of the absence of sublayers with differing stiffnesses, and the absence of rigid boundaries at the film surfaces, the wrinkles described in our experiment are homogeneous across the films.

\section{Experimental setup and materials}

\begin{figure}
\centering\includegraphics[width=0.8\columnwidth]{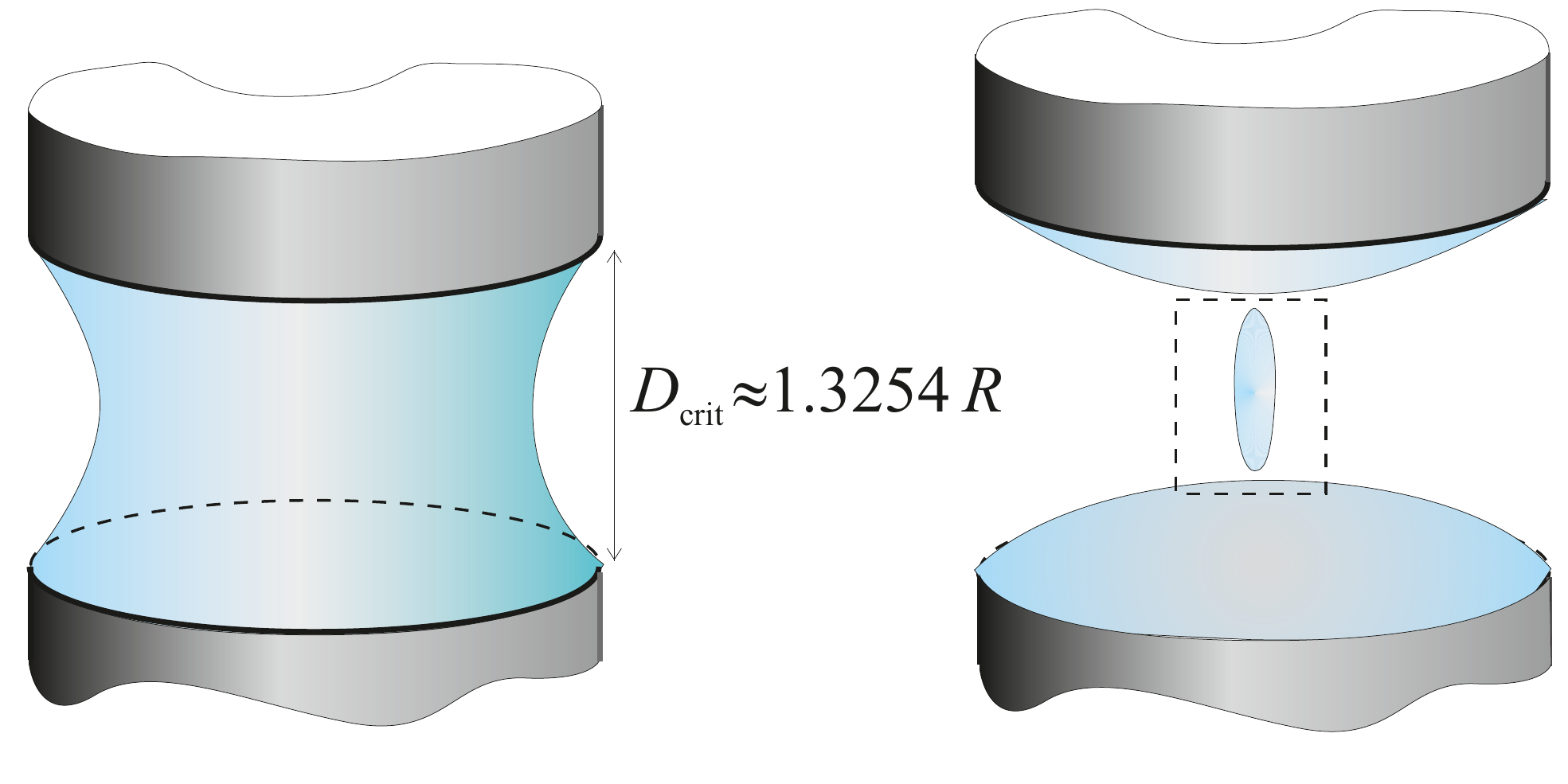}
\caption{Experimental geometry: two support rings and the catenoid before pinch-off (left) and the satellite bubble plus remnant films after pinch-off (right). The dashed rectangle sketches the maximum field of view of the side view camera. Most of the images shown in this paper are smaller clips of this field of view,
focusing on details.}
\label{fig:geo}
\end{figure}

Films were prepared from a room temperature smectic C mixture with 50 vol\% of 2-(4-n-Hexyloxyphenyl)-5-n-octylpyrimidine and 50 vol\% of 5-n-Decyl-2-(4-n-octyloxyphenyl) pyrimidine. The smectic C range is between 20 and 53$\,^\circ$\,C. We created the bubbles by a technique described earlier \cite{May2012},
sketched in Fig. \ref{fig:geo}:
The two coaxial circular rings are wetted with the smectic substance, then brought in contact. After their slow separation with a stepper motor, a catenoid shaped smectic film is formed. Since the catenoid exists as an
equilibrium film shape only up to a maximum ring separation of $D_{\rm crit}\approx 1.3254~R$ (with the inner
radius $R$ of the supporting rings), the catenoid collapses at this critical separation and leaves a freely floating, elongated bubble, as sketched in Fig.~\ref{fig:geo}.
With rings between 20 mm and 25 mm radius, the bubbles have radii of a few millimeters.
The catenoid collapse usually also leaves two circular films in the support rings which perform damped oscillations
to equilibrium.

Experiments were performed in microgravity during parabolic flights at NOVESPACE (Bordeaux)
with an Airbus 310, as well as in the ground lab under normal gravity. In both situations, bubbles form, but their properties differ in detail. Microgravity leads to a symmetric pinch-off of the upper and lower film parts, producing initially symmetric bubbles and a pair of films at the catenoid holders oscillating in antiphase. Those films periodically create air flow along the central axis that affects the bubble, squeezing
and stretching it.

In normal gravity, the pinch-off is asymmetric, the bubbles usually pinch off at their upper end first. The initial bubble shapes,
the relative phase of the film oscillations as well as the symmetry are affected. One problem with
experiments under normal gravity is that the bubbles tend to leave the observation window rather fast.

The bubbles were observed with two high-speed cameras, providing a side view and a bottom view. Within this study, exclusively the side views recorded with a Phantom v710 camera (operated at 5,000 to 10,000 fps) were evaluated.
We illuminated the bubbles with monochromatic parallel light, using a blue high-power LED with wavelength 460 nm.
The thickness of the smectic film was measured with a standard interference technique
developed for spherical smectic bubbles \citep{Stannarius1999}.
All films studied had submicrometer thicknesses, bubble diameters were in the millimeter to centimeter range.

Wavelengths of the wrinkles were determined from 2D Fourier Transforms of the optical images.
We selected for evaluation only film regions oriented normal to the viewing direction.

\section{Results and discussion}

\subsection{Shape dynamics}

As seen in Fig.~\ref{fig:1} and sketched in Fig.~\ref{fig:geo}, the bubbles have an elongated shape after pinch-off. They gradually
transform into spheres with minimal
surface during a period of several hundred milliseconds. This scenario has been analyzed in detail before \citep{May2012,May2014}.
On short time scales, however,
the smectic films behave like membranes with zero surface tension and fixed surface area.
During the shape transformations, the films permanently experience local compressions and dilations.
Air flow around the bubbles advects the smectic film material and permanently forces the films to reorganize locally.
The main reason for this air flow in our experiments
is the oscillation of the two remnant films at the support rings (Fig.~\ref{fig:geo}).
It depends in a complex fashion on the amplitude and phase of the two oscillating films.

The 2D flow in the smectic film by these friction forces can become locally divergent.
Positive divergence either leads to local film thinning (rupture of some layers),
or to the rupture of the whole film \citep{May2014}.
In smectic C films, film rupture may even induce a reduction of the soft tilt angle \citep{Trittel2017}.
Negative divergence compresses the film region.
When the surface area of the film cannot be reduced quickly enough by island formation \citep{May2014,Daehmlow2018}, the films need to buckle out of the plane.

In addition, bubbles that are far from the equilibrium sphere shape can undergo complex shape transformations.
They can, for example, develop invaginations at their ends, whereby a part of the bubble is pushed inwards (cf. Figs.~\ref{fig:1},\ref{fig:1a}).
In all these cases, the film may be subject to lateral compression. The response of the films is diverse and it depends upon
the speed of the compression and on details of the film structure: If the compression is sufficiently slow,
then the film remains flat and forms islands \citep{Daehmlow2018}. Fast lateral compression of a smectic film
cannot be compensated by this mechanism.
If the film is uniformly thick, or if it contains only very few islands, periodic wrinkling is the dominant response
that reduces the lateral extension without changing the film surface.
If, however, the film thickness is inhomogeneous because many islands already exist, then the film can respond primarily by a local out-of-plane bulging of these islands as described in the next section.
At higher compression rates, if bulging does not provide sufficient area reduction, it is accompanied by
wrinkling of the background film.

\subsection{Bulges and tubuli}

\begin{figure}
	\centering
	\includegraphics[width=1.0\columnwidth]{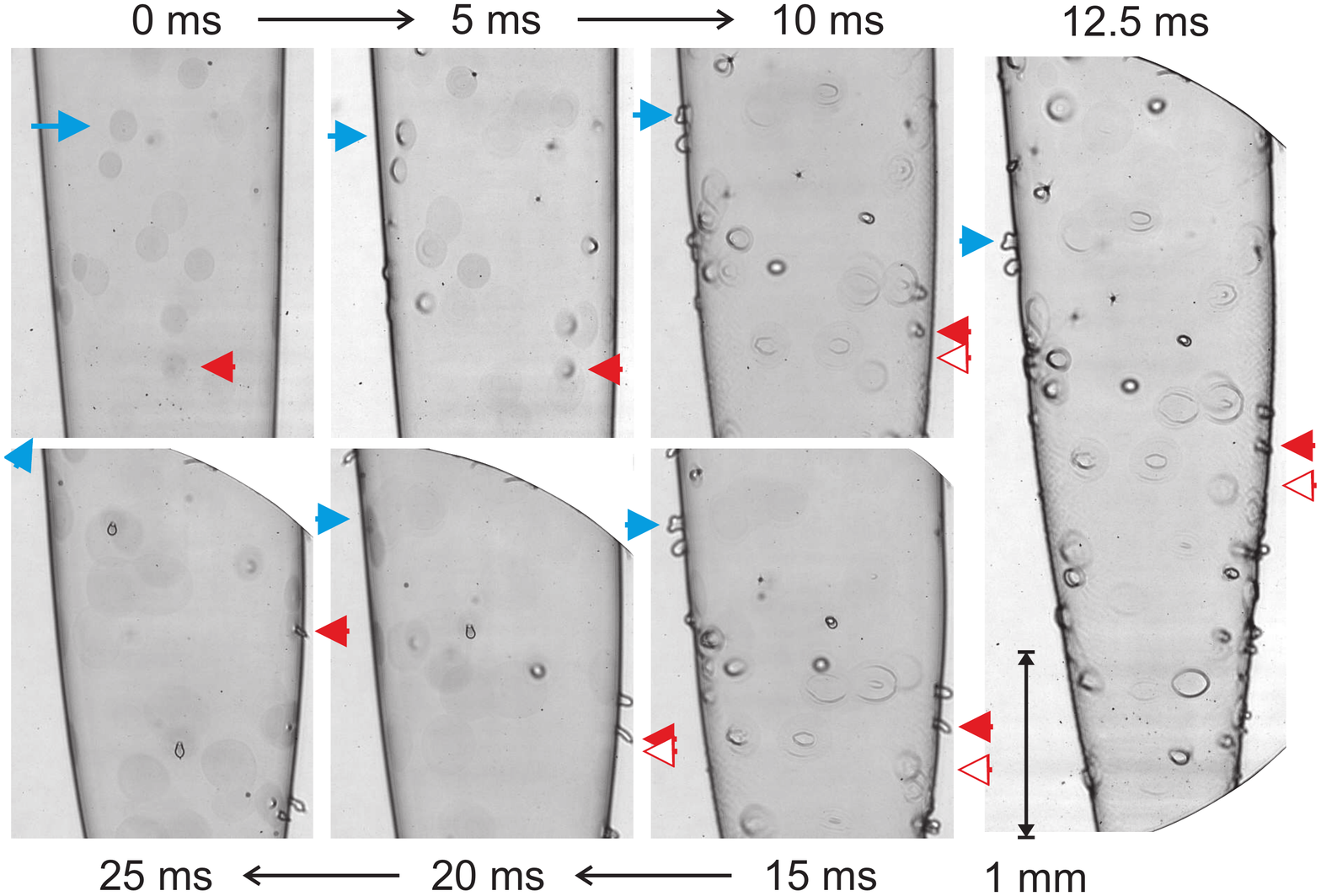}
		\caption{Formation of bulges during the lateral compression of a bubble consisting of an inhomogeneously thick film, observed in normal gravity. The film has a background
thickness of about 23 $\pm 3$ nm, islands of larger film thicknesses are seen as dark spots on the bubble.
Times are given respective to the first image. The 12.5 ms image shows a larger part of the bubble. Whitened corners in some images mask the regions outside the observation window. We note that the whole bubble slightly spins counterclockwise (when seen from above), thus the features on the front continuously move to the right. Explanation of the arrows see text.}
	\label{fig:bulges}
\end{figure}

Details of a bubble that is randomly speckled with islands of a few dozen micrometers diameter are shown in Fig.~\ref{fig:bulges}. These islands were formed previously during the collapse of a catenoid with inhomogeneous
film thickness.
This film reacts to the lateral compression with an extrusion of bulges at the centers of islands. Arrows point at two
selected features (islands/bulges) to visualize the axial contraction of the bubble. The open arrows indicate the
expected positions of the second feature if the film had not contracted.
The total axial contraction of the bubble in the displayed region amounts to about 10\%, with a maximum
shrinkage rate of about 2 - 3 \% per millisecond.
The time stamps given in the images refer to the first image in the sequence, which was taken 37 ms after pinch-off.

The islands represent 'germs', and bulges are formed in their centers (Fig.~\ref{fig:bulges}).
Even though it seems natural that any film thickness inhomogeneity breaks the translation symmetry and thus acts as a local generator of deformations, it appears counterintuitive at first glance that the extrusions appear in the thicker parts (islands) of the film: The energy
per area necessary to bend the smectic film is proportional to the film thickness (see next section), and islands are thus stiffer than the surrounding background film. Apparently
it should be easier to bend the thin background film rather than islands.
A potential reason for the observed phenomenon is the line tension created by the dislocations surrounding an island
\citep{Geminard1997}. It contributes an additional lateral compression force.
This also explains the experimental observation that smaller islands are more probable to form bulges.
The typical lateral extensions of the bulges are determined by the island sizes.
The maximum diameters of bulges are of the order of the island diameters.

The bulges form at a time scale of a few milliseconds, much faster than new islands can be generated.
Figure \ref{fig:bulges} shows a typical image sequence of a contracting section of a bubble (in the first 15 ms).
As long as the compression continues, the bulges grow and they can form tubuli which may even pinch off.
When the exposed region stretches again, bulges disappear.
This is seen in the second, bottom part of Fig.~\ref{fig:bulges}, for the following 10 milliseconds.

In flat film regions, the islands can bulge to both sides. Even the small excess pressure of a few Pascal inside the bubbles does not affect the preference direction of the bulges noticeably, they can grow inward and outward.
However, the local
mean curvature of the film breaks the symmetry and defines such a preference direction for bulging.
Bulges grow outwards in regions where the local bubble surface is convex, which is the common situation.
But they can also grow inwards where the local bubble surface is concave.
This is exemplarily seen in Fig. \ref{fig:1a}.

\subsection{Wrinkles}

The homogeneity of the films or the existence of islands depend upon the history. It is difficult to produce
catenoid films that are completely uniform. Film inhomogeneities anneal only if the film is given sufficient
time for equilibration (up to minutes). Then, one can obtain bubbles which initially contain little or no islands.

\begin{figure}
	\centering
	\includegraphics[width=1.0\columnwidth]{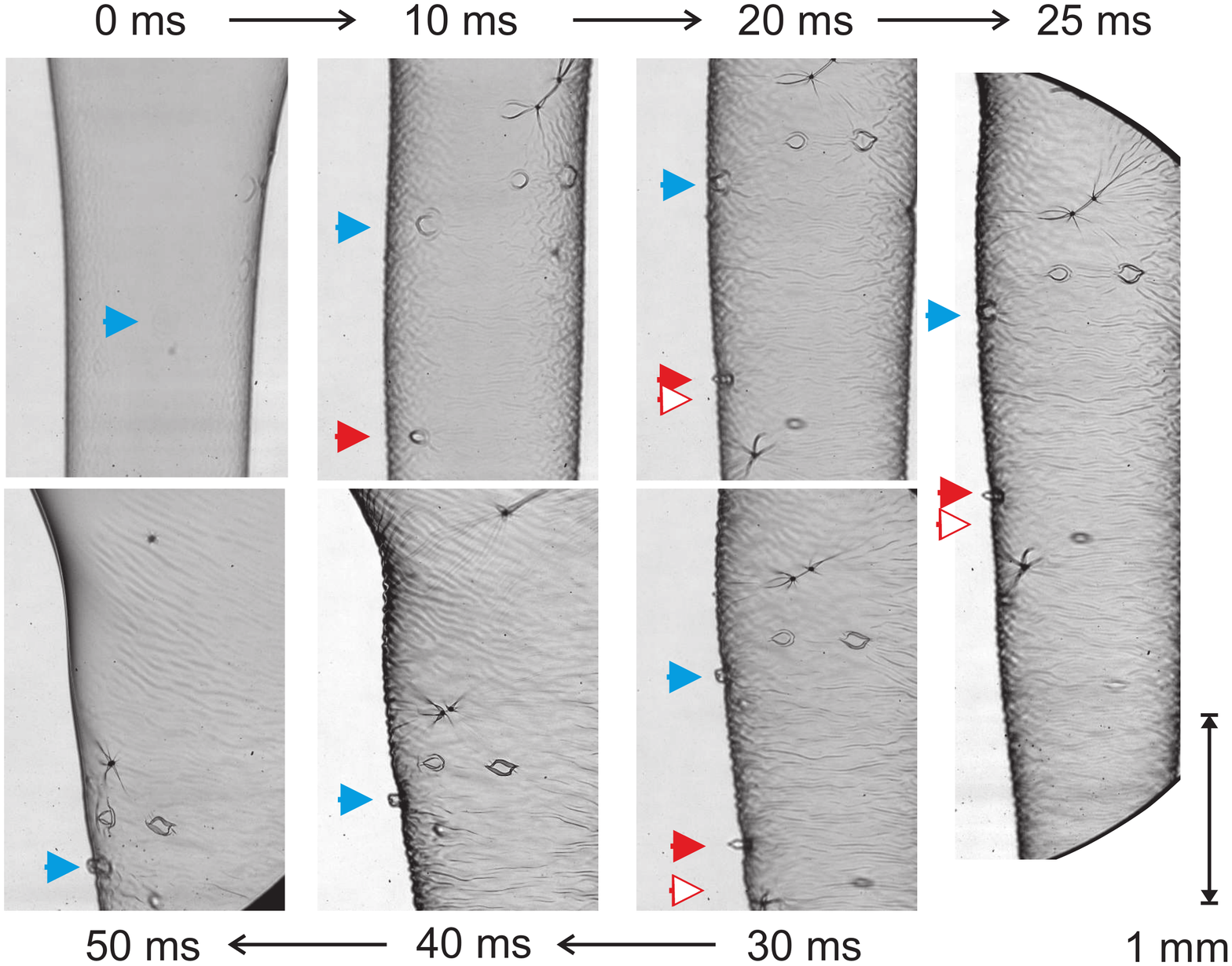}
		\caption{Formation of wrinkles during the lateral compression of an almost homogeneously thick film. The
bubble has a background film thickness of about 32 nm. The images show a selected area of the bubble surface.
Times are given respective to the first image. The 25 ms image shows a larger part of the bubble. Whitened corners mask the regions
outside the observation window. Arrows mark selected features, in order to collate film regions in subsequent images (see text and Fig.~\ref{fig:bulges}).}
	\label{fig:wrinkles}
\end{figure}

In bubbles of uniform film thickness, the film response to lateral compression is completely different from that
of inhomogeneous bubbles, as shown in Fig.~\ref{fig:wrinkles}.
The overall bubble geometry and shape dynamics in this example are very similar to those in Fig.~\ref{fig:bulges}, except that there are practically no islands on the film. This bubble section undergoes an axial contraction, and
wrinkles form, with a preferential orientation perpendicular to the contraction direction. The first frame shown
was taken 42.4 ms after pinch-off.
As in Fig.~\ref{fig:bulges}, arrows point at two selected features of the film to visualize the axial contraction of the bubble.
The open arrows indicate the expected positions of the second feature if the film had not contracted.
The axial contraction of the bubble in the displayed region amounts to about 10\%, with a maximum
shrinkage rate comparable to that of the bubble in Fig.~\ref{fig:bulges}.

\begin{figure}
	\centering
	\includegraphics[width=0.6\columnwidth]{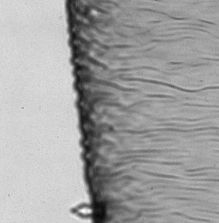}
		\caption{Detail of the wrinkle structure in side view (zoom of the 25 ms image in Fig.~\ref{fig:wrinkles}).}
	\label{fig:side}
\end{figure}

The film undulations can be directly seen in microscopic images at the sides of the bubble. Figure \ref{fig:side} shows a zoomed-in detail of the fully developed wrinkles.

When the contraction stress ceases, the wrinkles gradually disappear again (Fig.~\ref{fig:wrinkles}, bottom row). They can reappear
at a subsequent new contraction. During wrinkling, the surface area stays practically constant, and the orientation of the wrinkles
is directly coupled to the contracting direction.

\begin{figure}
	\centering
	\includegraphics[width=0.8\columnwidth]{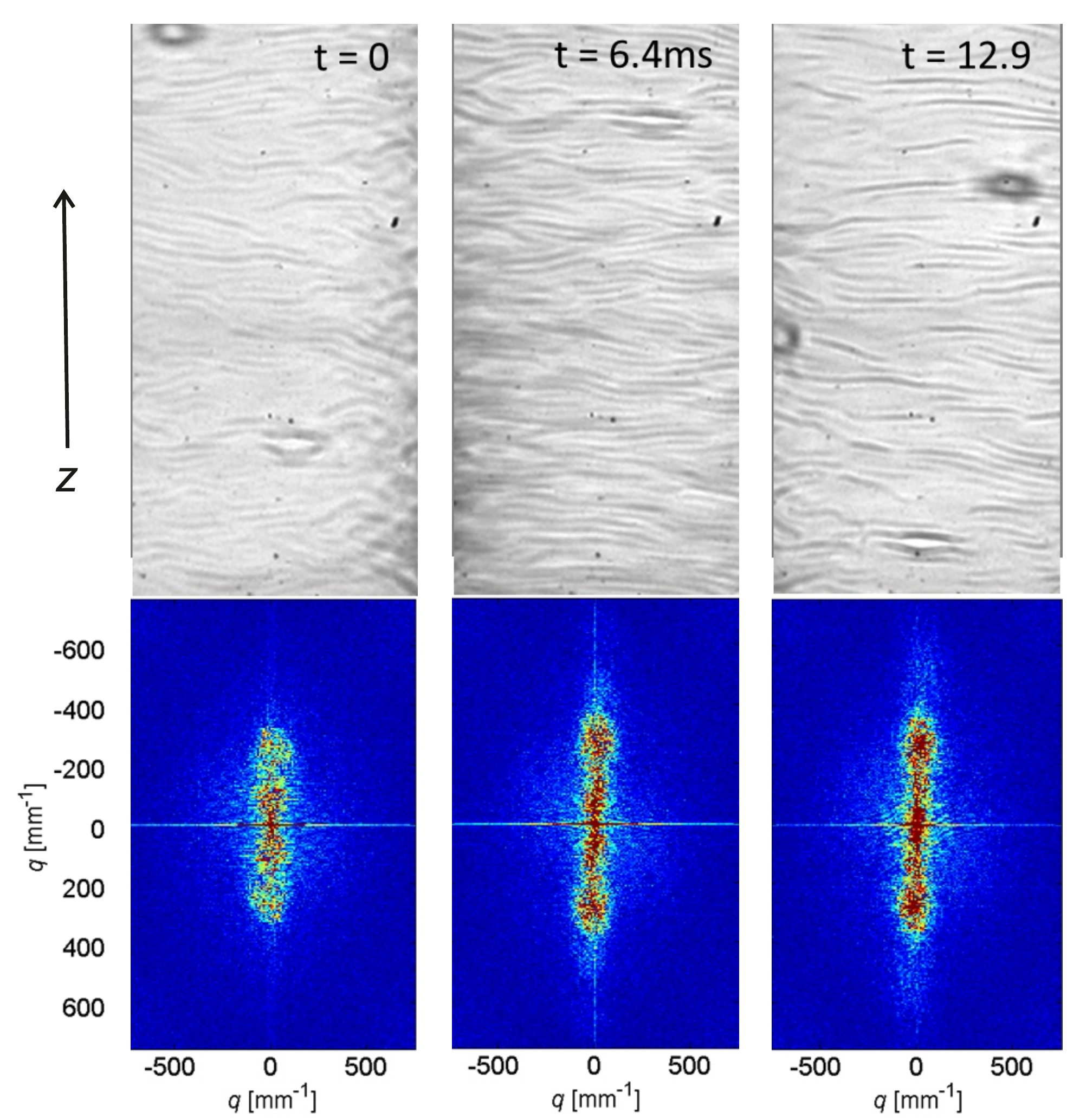}
	\includegraphics[width=0.7\columnwidth]{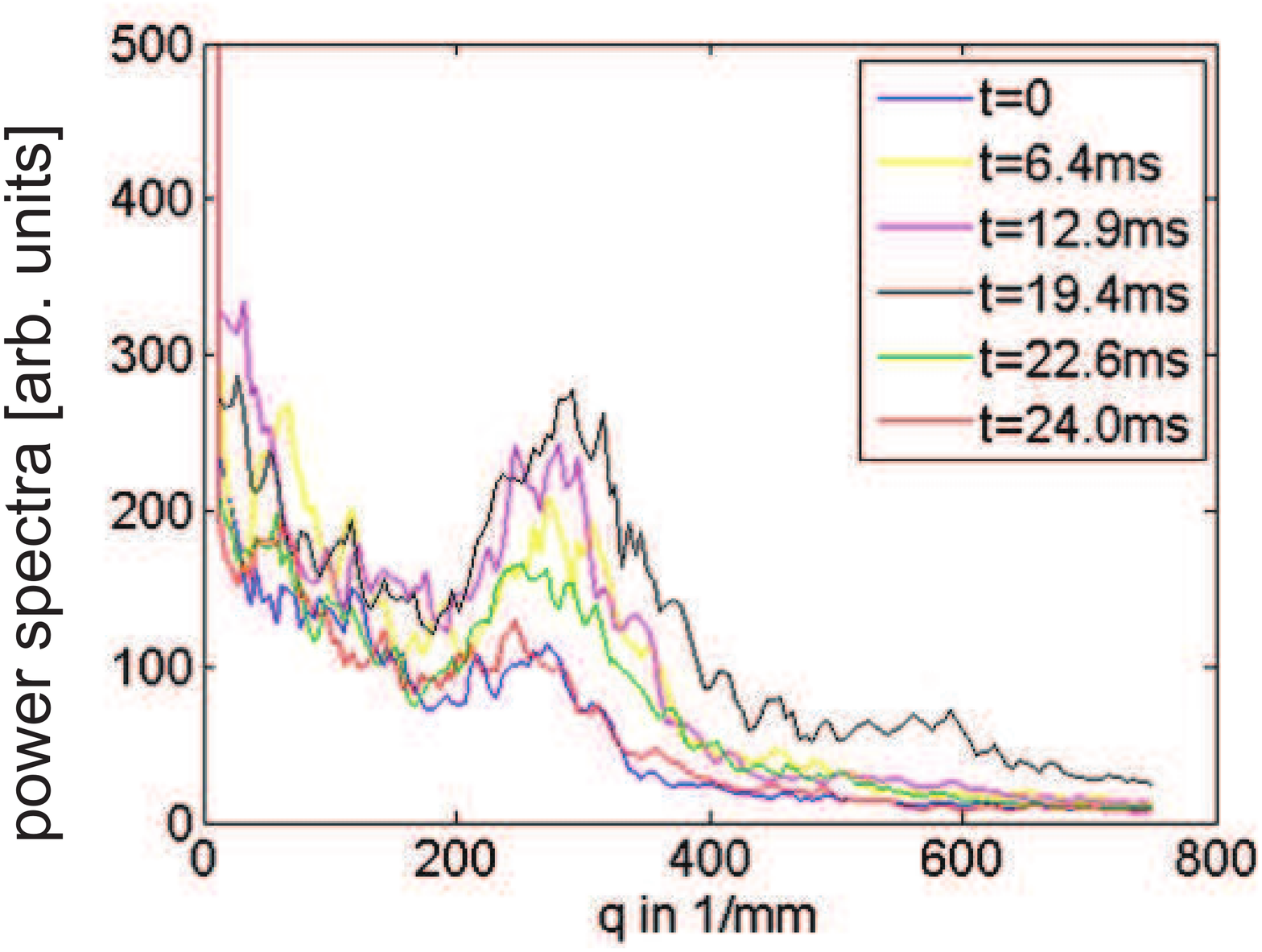}
	\caption{Wrinkles of an axially compressed bubble (top images), Fourier Transforms (middle) and vertical cross section profiles of the Fourier Transforms. The contraction stops after approximately 20 ms. Then,
the region expands again, the peak shifts back to lower wave numbers.}
	\label{fig:fft}
\end{figure}

Quantitatively, it is possible to measure the spectrum of the wrinkles by means of Fourier Transform (FT) of the optical images
of certain bubble regions. In order to avoid distortions, it is recommendable to choose regions where the viewing direction
is normal to the film. Figure \ref{fig:fft} shows three snapshots of wrinkles and the corresponding intensity of the 2D
FT power spectrum. It is obvious from the spectra that the wrinkles show a high degree of orientational ordering. From the
plot of the 2D cross sections of these spectra along the symmetry axis (vertical in the images) the graphs in the bottom image of
Fig.~\ref{fig:fft} were obtained. It is evident that the spectrum develops a characteristic peak around the wave number $q_{\max}\approx 300$/mm. This corresponds to wave lengths of about $20~\mu$m of the wrinkles. Furthermore it is evident that the wave length decreases
with time. This is the consequence of the material-fixed position of the wrinkles. The number of wrinkles in a certain film region
remains approximately constant, and the maximum of the spectrum shifts to shorter wavelengths (larger $q$) as a consequence of the lateral
compression of the structure.

\begin{figure}
 \center  \includegraphics[width=1.0\columnwidth]{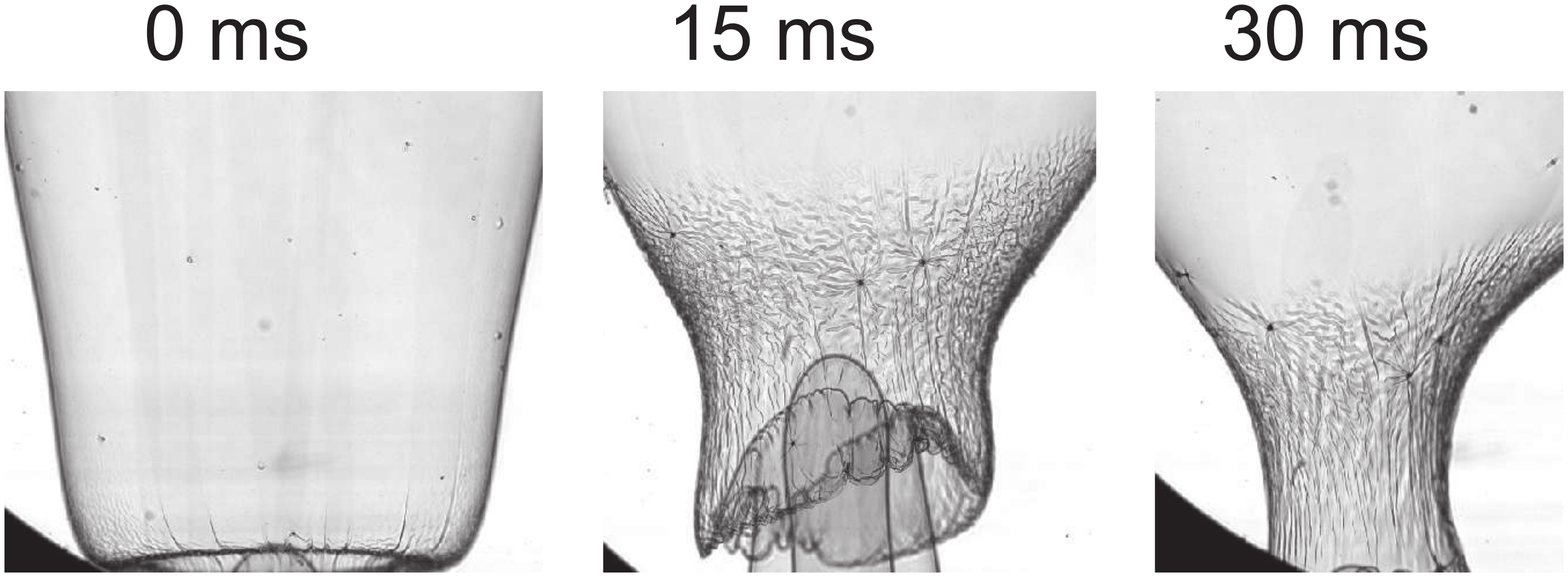}
	\caption{Local contraction of the diameter of the freely floating bubble leads to axial wrinkles.
The dark corners are outside the observation window. Image heights are 2~mm. }
	\label{fig:quer}
\end{figure}

Another scenario where wrinkles can form is the radial contraction of a tubular bubble section, as seen in Fig.~\ref{fig:quer}.
Here, the wrinkles are generated with a preferential axial direction respective to the bubble.
In that geometry, the compression arises from
the local shrinkage of the bubble circumference.
Typical wavelengths appear to be slightly larger than in the former, tangential patterns.
Often, the shape transformations of bubbles involve more complex scenarios, where both the axial contraction
and the diameter shrinkage occur concurrently. In that case, less ordered patterns are observed. A typical example
is that of Fig.~\ref{fig:1a} where wavelengths with different orientations are superimposed.

\begin{figure}
	\centering
	\includegraphics[width=1.0\columnwidth]{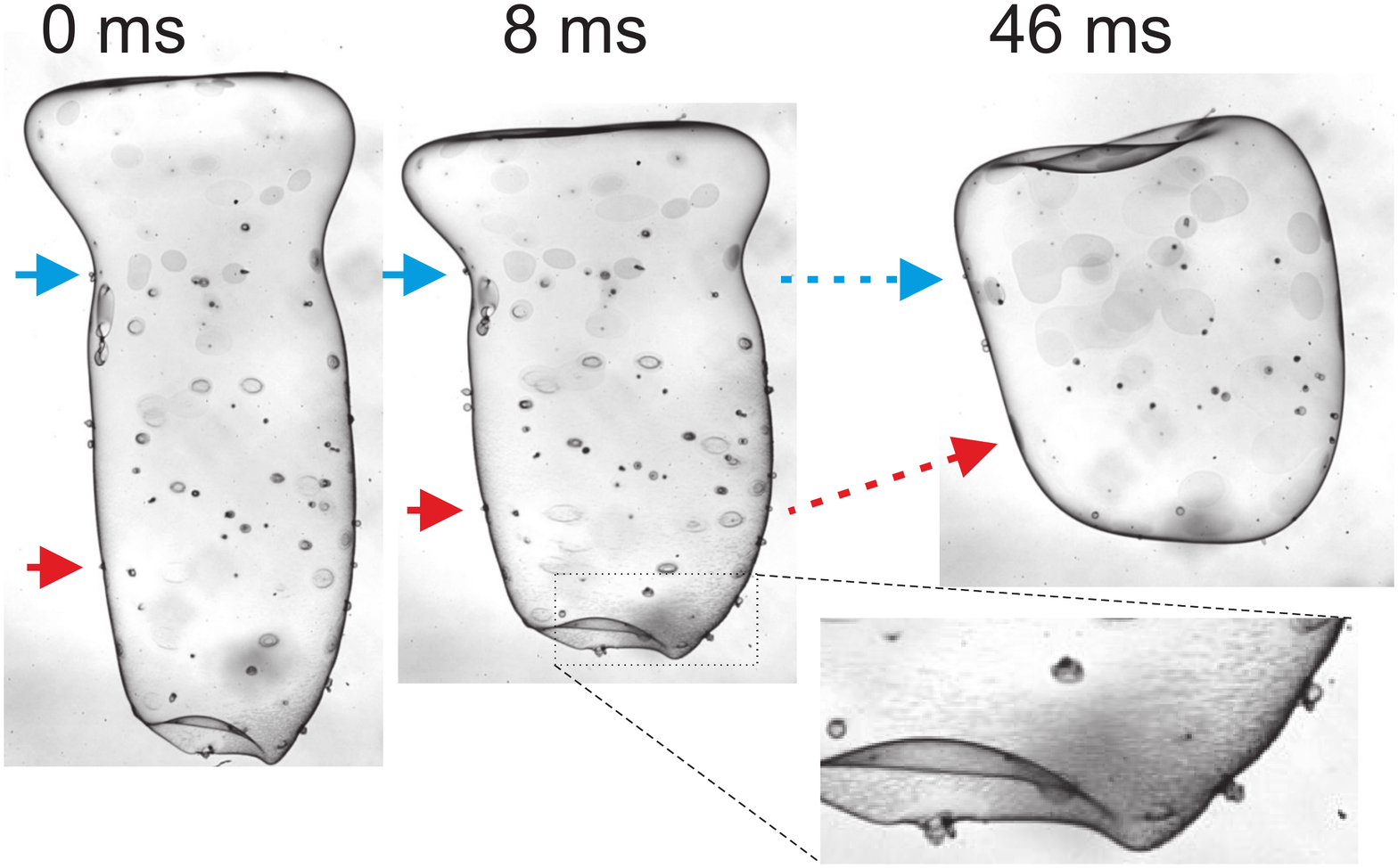}
		\caption{Bubble with a background film thickness of 16 nm, observed in microgravity:
 Bulges form only in the contracting parts. In the second image, taken 8 ms after the first one, shallow wrinkles form at the bottom end where the axial compression rate is strongest. The bottom image is a zoomed view of this region.
 After another 38 ms, the wrinkles disappeared and the bulges gradually shrink (right image). }
	\label{fig:bulges2}
\end{figure}
Figure \ref{fig:bulges2} shows a bubble of more exotic shape during a phase when it is compressed in the central and lower parts while being only little deformed in the upper part. Here, we find the combination of both deformations,
bulges and wrinkles.
Even though islands are uniformly distributed across the surface of this bubble, bulges are formed almost exclusively in the contracting central and lower regions. Islands in the top part, less exposed to lateral film area contraction, remain flat.
In the bottom part of the bubble, the axial compression is so strong that it cannot be compensated by
bulging. Therefore, wrinkles are created in addition. The third image was taken during a phase where the
film shape reorganization has slowed down considerably, the wrinkles have vanished and the bulges gradually disappear.

\section{Dynamic wrinkling model}

The model developed in this section is intended to describe the simple structures shown in Figs.~\ref{fig:wrinkles}, to \ref{fig:bulges2}.
The complex air flow plays a decisive role in the present system.
In order to demonstrate the principal mechanism without providing a deeper quantitative analysis, we base our model on some crude approximations.

\begin{figure}
\centering\includegraphics[width=0.5\columnwidth]{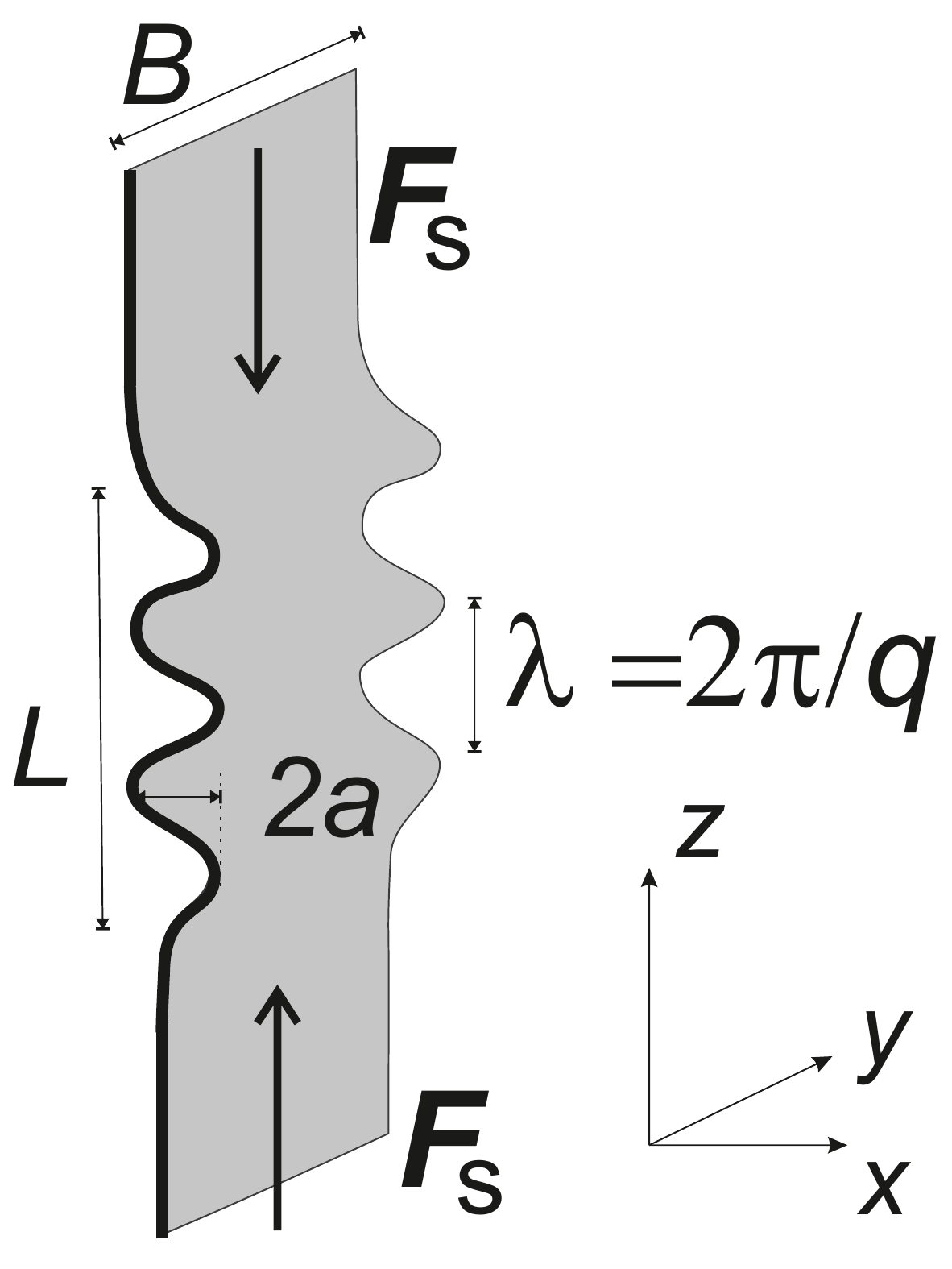}
\caption{Film geometry and definition of the coordinates.}
\label{fig:wrinklegeo}
\end{figure}

The model considers a rectangular film area of a given fixed width $B$ and initial length $L_0=L(t=0)$. This area be compressed uniaxially
by opposing forces $\vec F_s$ along $z$ and $-z$, resp. (see Figure \ref{fig:wrinklegeo}).
The detailed nature of these compression forces is not relevant for the pattern selection.
In the axially contracting bubbles,
it is generated by pressure acting on the upper and lower caps of the bubble and by friction of the
axially streaming surrounding air. The response of the
film is the formation of $N$ wrinkles with initial wavelength $\lambda_0=L_0/N$ and amplitude $a(t)$.
Small deformations are approximated by the harmonic ansatz $X(z)=a\sin q z$, with $q=2\pi/\lambda$.
The initial wavelength is
$q_0=2\pi N/L_0$. During the compression, the number of wrinkles remains roughly conserved (wrinkles are material-fixed)
and their wavelength decreases. Since we are interested in the initial wavelength selection process, we can safely
consider
the wavelengths to be approximately constant, $q\approx q_0$, the temporal variation does not exceed $\approx$10\%.

The driving force per width of the rippled film region is $f_s=F_s/B$.
This force performs a mechanical work $W_s$, which is in part stored in the deformation of the smectic
film ($W_{\rm b}$). The other part, $W_{\rm f}$, is dissipated primarily by the redisplacement of air near the
 film during the growth of the undulation amplitude.
We thus set $W_{\rm f}=W_{\rm s}-W_{\rm b}$.
The mechanical work of the driving external forces performed on the film during compression is
$W_s(t) = f_s B (L_0-L(t))$.
From the relation $L_0=\int_0^L {\sqrt{1+(dX/dz)^2}} dz$, one finds $L_0\approx (1+a^2q^2/4)L$,
thus
$$ W_{\rm s} = \frac{f_s B L}{4} a^2 q^2 .$$
The reduction of the lateral film area (at given amplitudes $a$ of the undulation) is more effective
for shorter waves, this will provide a suppression of the long-wavelength tail of the
unstable mode band.

Bending of smectic membranes is related to a free energy density contribution of the form
\begin{equation}
\label{eq2}
w_{\rm b}  = \frac{1}{2} K (X'')^2.
\end{equation}
The main contribution to this term is the splay deformation of the director $\vec n(z)$,
\begin{equation}
\label{eq2a}
w_{\rm b}  \approx w_{\rm elast}  = \frac{1}{2} K_{11} (\nabla \vec n)^2
\end{equation}
which in the
smectic A phase is linearly coupled to layer undulations
$$ \quad (\nabla \vec n) =\frac{d n_z}{d z}=-\frac{d^2 X}{d z^2}$$
in our geometry.
Other potential contributions to the smectic layer bending elasticity \cite{Santangelo2005,Fujii2011} might add to that. For example,
the layer compression modulus may lead to some minor corrections, but the splay term remains to be the dominant one.
Integrated over the wrinkled region, one obtains (with the averages $\overline{\sin^2 qz}=\overline{\cos^2 qz}=1/2$)
\begin{equation}
\label{eq2b}
W_{\rm b}  = \frac{K L B h}{4}  a^2 q^4 ,
\end{equation}
where $h$ is the film thickness.
This model is strictly correct only for a smectic A film, and the elastic deformations are more complicated in the smectic C bubbles
studied here. But since we are interested primarily in the qualitative elucidation of the instability mechanism, this detail is
of secondary importance. The functional dependence on $a$ and $q$ in Eq.~(\ref{eq2b}) will not be different in our smectic C films.
Experiments with wrinkling smectic A bubbles would be advantageous for the modeling, but earlier experiments \cite{May2014} proved that it is much more difficult to obtain oscillating smectic A bubbles. They are much more vulnerable to bursting than the smectic C films investigated in our experiments.

Now, we need to relate the growth rate of the undulation to the flow of air from the crests to the valleys of the undulation wave, and consequently to the pressure profile $p(z)$ along the film.
We adapt a concept used in earlier descriptions of transient pattern formation in thin films,
 e. g. \cite{Vrij1968}, and relate the pressure gradient profile with the air flow $j_z$ along the film
($z$ direction) by the equation
\begin{equation}
j_z = -C p'(z),
\end{equation}
where $C$ is a geometrical constant that describes the complex flow processes redistributing
the air along the bulging film on both sides. The quantitative determination of this constant is not straightforward,
but one can make the reasonable assumption that it is roughly the same for all wavelengths.
The related pressure modulation has the same periodicity
as the film deflection, $p=p_0 \sin q z$.
The local film deformation changes according to
\begin{equation}
\dot X = \frac{d j_z}{d z}=  - C p''(z) .
\end{equation}
This yields the relation between the growth of the deformation amplitude and the pressure
amplitude
\begin{equation}
\dot a = Cq^2 p_0,  \quad p= \dot a  C^{-1} q^{-2} \sin qz .
\label{eq:4}
\end{equation}
The energy dissipated in this process is
\begin{equation}
W_{\rm f}= \int_0^L p dV = B\int_0^L p X dz .
\label{eq:0}
\end{equation}

Inserting Eq.~(\ref{eq:4}) into the equation~(\ref{eq:0}) for the dissipated energy $W_{\rm f}$, we find
\begin{equation}
W_{\rm f}=  B\int_0^L C^{-1} q^{-2} \sin^2(qz) a \dot a \,dz,
\end{equation}
and averaging $\overline{\sin^2 qz}=1/2$ yields
\begin{equation}
W_{\rm f}=  \frac{BL}{2C q^2} a\dot{a}.
\end{equation}
We arrive
at the dispersion relation
\begin{equation}
\dot a = {(f_s q^2-K h q^4)\cdot \frac{C}{2} q^2}a= (f_s q^4- K h q^6)\frac{C}{2}a,
\end{equation}
which describes exponential growth of an unstable mode band between $q=0$ and $q_c = \sqrt{f_s/(Kh)}$.
Short wavelength wrinkles are suppressed by the bending elasticity of the smectic layers, large wavelength wrinkles
grow very slowly, thus there exists an optimum growth rate for an intermediate wave length.
The fastest growing mode is
\begin{equation}
q_{\max} = \sqrt{\frac{2}{3} \frac{f_s}{Kh}},
\end{equation}
with the maximum growth rate
\begin{equation}
s_{\max}=\frac{2C}{27}\frac{f_s^3}{K^2h^2}.
\label{smax}
\end{equation}
Thinner films develop wrinkles faster than thicker ones.
An estimate of $f_s$, with available material parameters and experimental data,
$K\approx 10$~pN, $h\approx 20$~nm, and $q_{\max}=300$~mm$^{-1}$, yields reasonable 3~nN/m.
For quantitative predictions on the growth rates, one would need a guess of the geometrical constant $C$.
On the basis of the experimental data and Eq.~(\ref{smax}), we arrive at an estimate of $C\approx 10^{-3}$m$^5$~N$^{-1}$s$^{-1}$ .

We summarize these results as follows:
A lateral compression of the film that is faster than the time scale necessary
to reorganize the film thickness (i. e. island growth) leads to a growth of ripples within a certain
mode band. The wrinkles are aligned with the wave vector along the compression forces.
The fastest growing mode depends upon the bending rigidity of the film, the wave
number becomes smaller with larger $K$. The wave number also decreases with film
thickness, because the energy needed to buckle the film linearly increases with
film thickness. 
An in-plane compressing force $f_s$ is the essential ingredient
for the development of the wrinkling instability. It needs to exceed a threshold
set by $K h$ in order to achieve the formation of an unstable mode band with sufficiently quick
growth rates. The larger the compression force $f_s$, the shorter wavelengths of the wrinkles are expected.
The forces $F_s$ can arise from pressure onto the upper and lower caps of the elongated bubbles, from
friction of a convergent air flow near the film surface, or axial contraction of the bubble (Fig.~\ref{fig:quer}).
Their quantitative determination is complicated and was not attempted within this study.
The film surface is unchanged during wrinkling.
The undulations are material-fixed, thus the dominant wavelength shortens during contraction proportional to $L/L_0$.
When the area stretches again, the wavelength grows again and the pattern vanishes.

\section{Conclusions and summary}

We have demonstrated that smectic freely floating bubbles can exhibit a novel type of spontaneously formed dissipative pattern,
the wrinkling of thin films. It occurs as a transient instability when elongated bubbles are compressed axially
or when a tube-like bubble section contracts. The patterns form only when the films are exposed to strong, fast compression
stresses in the film plane. The film surface area remains approximately unchanged. When the compression stresses cease or
transform into dilation, the wrinkles and bulges disappear again. Typical timescales for growth and disappearance are few milliseconds.
Typical wavelengths are in the range of a few dozen micrometers. It is obvious that this phenomenon cannot be observed in conventional soap
films which can respond to lateral compression by practically instantaneous thickness growth and film area reduction. The observed
phenomenon is a peculiarity of the smectic structure.

The second phenomenon described here is the extrusion of bulges from the plane in films that contain islands. Here, the analysis
of the instability mechanism in detail and the description of the quantitative relations between island sizes,
line tensions, external compression forces and bulge sizes still remains a challenge.
A better spatial resolution in the experiment will be required for that.

A key ingredient of our experiment are the oscillating drum-like smectic membranes that remain at the catenoid holders and that
produce strong air flows in axial direction, distorting the freely floating bubbles. More quantitative experimental data might be
obtained when the bubbles are exposed to well-controlled acoustic excitation fields. This could be a challenging future task.
The obvious reason that dynamic wrinkles have not been described before is that practically all earlier investigations of smectics,
with few exceptions, have been performed either in sandwich cells with rigid interfaces or in thin free-standing films spanned across
fixed supports. A desirable improvement of the experimental investigation of wrinkles of smectic films would be the study of
laterally compressed planar films. Several technical problems have to be solved for such an experiment. It would allow to
compare smectic C and smectic A phases using polarizing microscopy, and to identify the coupling of
the c-director dynamics to layer undulations.

Other structured fluid films like lipid membranes or Newton black films should exhibit a similar instability when exposed to sufficiently fast
lateral compressions.
Apart from the general relevance of this unique phenomenon in freely suspended fluid films, the analysis of wrinkles may provide
quantitative access to material parameters of smectic LCs that are otherwise very difficult to measure.

\section*{Acknowledgments}

This study was performed with financial support from DLR within Project OASIS-Co (50WM1430 and 50WM1744) and from DFG (STA 425/40-1). Participation of Robert G\"opfert and Vincent Nathow in the ground lab measurements is gratefully acknowledged.

\bibliographystyle{unsrt}

\end{document}